\documentclass[
    ,final            
  ]
  {aipproc}

\layoutstyle{6x9}

\begin{document}

\title[Observational constraints on BH growth]{Where the Wild Things
  Are: Observational Constraints on Black Holes' Growth}

\classification{98.54.-h; 98.80.Es}
\keywords      {quasars; galaxy formation; cosmology.}

\author{Andrea Merloni}{
  address={Excellence Cluster Universe, TUM, Boltzmannstr. 2,
  85748, Garching, Germany \& Max-Planck-Institut f\"ur
  Extraterrestrische Physik, Giessenbachstrasse, D-85741 Garching, Germany}, 
, email={am@mpe.mpg.de}}

\begin{abstract}
The physical and evolutionary relation between growing supermassive
black holes (AGN) and host galaxies is currently the subject of
intense research activity. Nevertheless, a deep theoretical 
understanding of such a
relation is hampered by the unique multi-scale nature of the combined
AGN-galaxy system,  which defies any 
purely numerical, or semi-analytic approach. 
Various physical process active on
different scales have signatures in different parts of
the electromagnetic spectrum; thus, observations at different wavelengths and
theoretical ideas all should contribute towards a ``large dynamic range'' view
of the AGN phenomenon.
As an example, I will focus in this review 
on two major recent observational results on the cosmic
evolution of supermassive black holes, 
focusing on the novel contribution given to the
field by the COSMOS survey. First of all, I will discuss
the evidence for the so-called ``downsizing'' 
in the AGN population as derived from
large X-ray surveys. I will then present new constraints on the evolution of
the black hole-galaxy scaling relation at $1<z<2$ derived by exploiting
the full multi-wavelength coverage of the survey on a complete
sample of $\sim$90 type 1 AGN.
\end{abstract}

\maketitle


\section{Introduction}

In the past decade three seminal discoveries have revealed tight links
and feedback loops between the growth of nuclear super-massive black
holes and galaxy evolution, promoting a true
 shift of paradigm in our view of black hole astrophysics, which
have moved from the role of exotic tracers (the `Wild Things')
of cosmic structures to
that of fundamental ingredients of them.

First of all, the search for the local QSO relics via the study of
their dynamical influence on the surrounding stars and gas
 led to the discovery of
SMBH in the center of most nearby bulge-dominated galaxies. The steep
and tight correlations between their masses and bulge properties
(so-called {\it scaling relations};
\cite{gebhardt:00,ferrarese:00,marconi:03}) represented the first and
fundamental piece of evidence in favor of a connection between galaxy
evolution and central black holes.
The second one stems from the fact that SMBH growth is now known to be
due to radiatively efficient accretion over cosmological times, taking
place during ``active'' phases  \citep{marconi:04,merloni:08}
(hereafter MH08). If most galaxies host a SMBH today, they should have
experienced such a phase of strong nuclear activity in the past.
Finally, extensive programs of optical and NIR follow-up observations
of X-ray selected AGN in the {\it Chandra} and {\it XMM-Newton} era
 put on solid grounds the evolution of accretion
luminosity over a significant fraction of cosmic time. We have thus
discovered that lower luminosity AGN peak at a lower
redshift than luminous QSOs \citep[see e.g.][]{hasinger:05}. Such a
 behavior is analogous to that observed for star
formation (usually referred to as ``cosmic downsizing'') lending a
third, independent, support to the idea that the formation and
evolution of SMBHs and their host galaxies might be closely related.

Such a shift of paradigm has sparkled the activity of theoretical
modelers. Following early pioneering analytic works
\citep{silk:98,fabian:99}, widely different approaches have been taken to
study the role of AGN in galaxy evolution. Semi-analytic models (SAM)
have been the more numerous \citep[see e.g.][]{croton:06}, and, 
despite their huge
freedom in the exploration of parameter spaces, have helped
establishing the importance of late-time feedback from radio active
AGN for the high-mass end of the galaxy mass function. On the other hand, fully
hydrodynamic simulations of cosmological BH have been performed \citep[see
e.g.][]{dimatteo:08}, but their computational costs have so far
allowed only a limited exploration of sub-grid prescriptions. 
A third, hybrid, approach has also been followed, in
which the results of high-resolution simulations of galaxy-galaxy
mergers with black holes \citep{dimatteo:05} have been used to
construct a general framework for merger-induced AGN feedback
\citep{hopkins:06}.
The combination of these approaches has sharpened our
view of the SMBH-galaxy co-evolution, but none of them can be
considered self-sufficient in such a heavily observation-driven field
of astrophysics. The reason for this bottleneck is the unique
multi-scale nature of the problem at hand, which defies any current
(and foreseeable) purely numerical, or semi-analytic approach. 

\begin{figure}
  \includegraphics[height=.35\textheight]{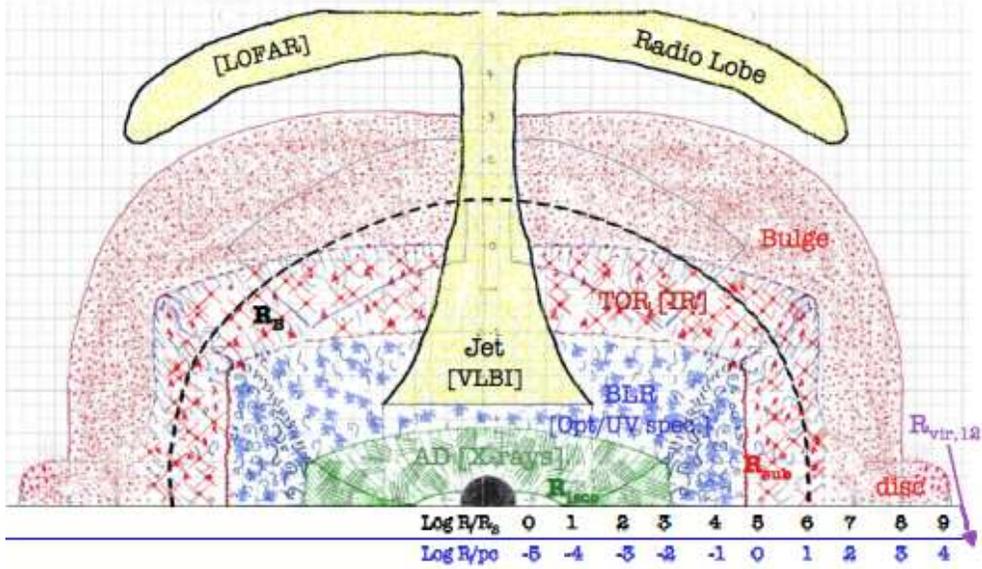}
  \caption{A schematic, logarithmic view of an AGN-galaxy
    system. Scales on the bottom axes are in units of Schwarzschild
    radii and parsec, and are approximately inferred for a $\sim 10^{11}
    M_{\odot}$ galaxy containing a $\sim 10^8 M_{\odot}$ black
    hole. Here, $R_{\rm vir,12}$ is the virial radius of a $10^{12}
    M_{\odot}$ dark matter halo, $R_{\rm B}$ stands for the Bondi
    radius (marked also by a thick dashed line), $R_{\rm sub}$ and
    $R_{\rm ISCO}$ for the dust sublimation radius and the innermost
    stable circular orbit ($\sim$inner edge of the accretion disc),
    respectively. The logarithmic scales at the bottom are in units of
    Schwarzschild radii ($R_{\rm S}$) and parsec, respectively.
   See text for details.}
   \label{fig:log}
\end{figure}
 
Let us illustrate this point schematically. An AGN releases most of
its energy (radiative or kinetic) on the scale of a few Schwarzschild
radii ($\sim 10^{-5}$ pc for a $10^8$ $M_{\odot}$ BH), while the mass inflow rate
(accretion rate) is set at the Bondi radius, some $10^4$ times further
out; broad permitted atomic emission lines, used to estimate SMBH
masses in QSOs, are produced at $\sim$0.1 pc; on a scale of a parsec, and
outside the sublimation radius, a dusty, large-scaleheight, possibly
clumpy, medium obscures the view of the inner engine \citep{elitzur:08}
crucially determining the observational properties of the AGN
\citep{netzer:08}; on the same scale, powerful star formation might be triggered
by the self-gravitational instability of the inflowing gas; finally,
AGN feedback must be operative on the galaxy scale ($\sim$ a few kpc, some
$\sim 10^8$ times $R_S$!) in order to have an appreciable effect  on its global
structural properties. A schematic logarithmic map of a AGN-galaxy
system is  shown in Fig.~\ref{fig:log}, where I have highlighted the
various physical regions of crucial interest and the wavelength ranges
where the associated emission takes mostly place.
Bringing all of the above into a coherent framework is indeed a
formidable challenge. In fact, each physical process active on
each different physical scale has a signature in a different part of
the electromagnetic spectrum (so that different instruments are needed
to unveil it). Observations at different wavelengths and
theoretical ideas all contribute towards a ``large dynamic range'' view
of the AGN phenomenon, capable of conceptually ``resolving'' the many
physical scales involved.

In this review, I'd like to show two specific examples of how deep,
multi-wavelength surveys can be used to deepen our knowledge of the
AGN phenomenon, of its evolution over cosmic time, and of its
relationship with the evolution of the galaxies they are embedded in.
In particular, most of the recent results shown here are part of the
large collaborative effort known as the COSMOS survey ({\tt
  http://cosmos.astro.caltech.edu/}).

\section{Wild things in the COSMOS}
The Cosmic Evolution Survey (COSMOS) field
\citep{scoville:07} is a so far unique
area for deep and wide comprehensive multiwavelength coverage,
from the optical band with {\it HST, Subaru} and other ground
based telescopes, to the IR ({\it Spitzer}), and X-rays with {\it
  XMM-Newton} and 
{\it Chandra}, to the radio with the {\it VLA}. The spectroscopic
coverage with {\it VIMOS/VLT} (zCOSMOS; \citep{lilly:07}) and {\it IMACS/Magellan}, coupled
with the reliable photometric redshifts derived from multi-band
fitting allow us to build a large and homogeneous sample of both
obscured and un-obscured AGN
 with dense spectral coverage, to estimate the effects of
intrinsic reddening and to keep under control selection effects.

I have chosen to focus here on just two among the many recent COSMOS
results in the field of AGN, in order to illustrate the power and
versatility of this kind of surveys. First of all, I will show how the
very high level of (spectroscopic plus photometric) redshift
completeness of the X-ray selected AGN samples allows an exquisite
determination of the AGN luminosity function evolution, and how,
coupling such a study with basic theoretical consideration on how black
holes evolve over cosmological times (via a continuity equation), we
can study in great details the downsizing phenomenon for AGN.
Second, I
will briefly mention recent progresses made on the study of the
cosmological evolution of the scaling relations between black holes
and host galaxies made possible by the intensive multi-wavelength
coverage of the COSMOS field.

\subsection{Dissecting AGN downsizing}
The term {\it downsizing} was first used by \citet{cowie:96} to
describe their finding that actively star-forming galaxies at low
redshift have smaller masses than actively star-forming galaxies at
$z\sim 1$. In the
current cosmology jargon, this term has come to identify a
variety of possibly distinct phenomena, not just
related to the epoch of star formation, but also to that of star
formation quenching, or galaxy assembly (see the discussion in \citet{faber:07},
and references therein).

\begin{figure}
\begin{tabular}{ll}
  \includegraphics[height=.27\textheight]{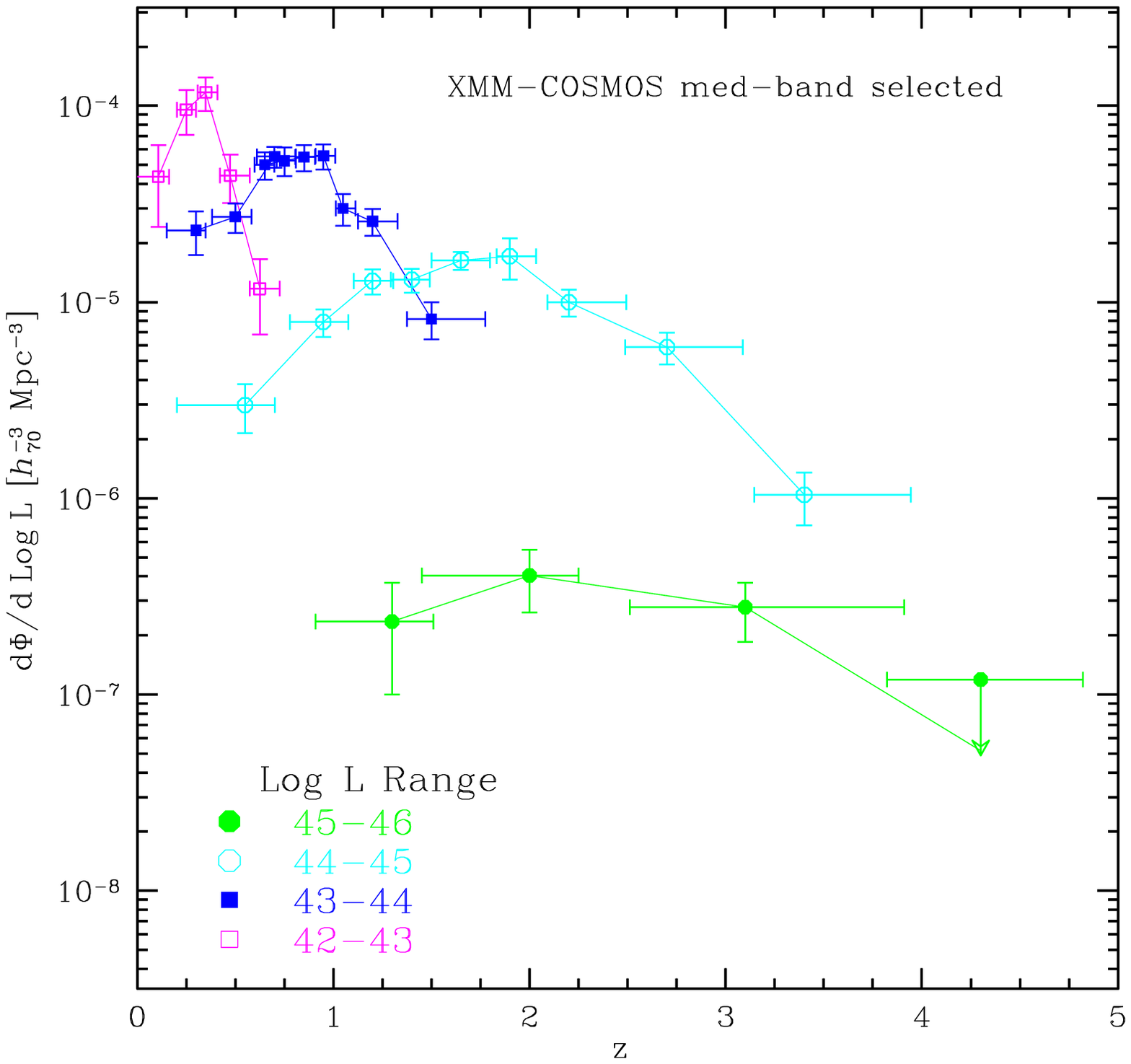}&
  \includegraphics[height=.30\textheight]{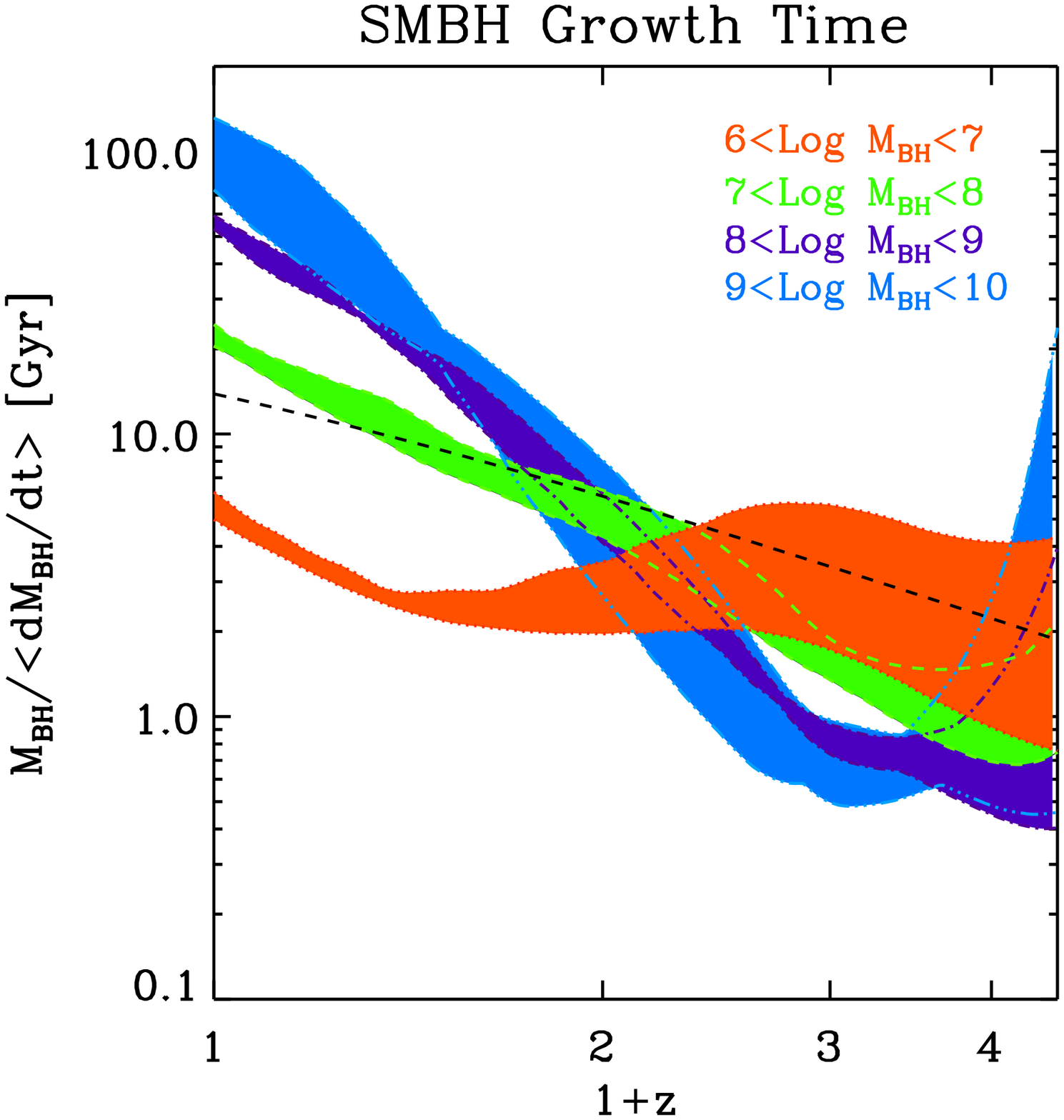}\\
\end{tabular}
  \caption{{\bf Left}: Number density of 2-10 keV X-ray selected AGN in the
    XMM-COSMOS field as a function of redshift for different
    luminosity bins. AGN phenomenological downsizing is evident in
    the difference between the epochs of peak activity of objects of
    different luminosity (Miyaji et al., in preparation). {\bf Right}: Average Growth time of Supermassive Black Holes (in years)
    as a function of redshift for different black hole mass ranges. 
    The dashed line marks the age of the Universe; 
    only black holes with instantaneous
    growth time smaller than the age of the Universe at any particular
    redshift can be said to be effectively growing.}
   \label{fig:xmm}
\end{figure}

 Given this growing body of observational
 evidence, it is legitimate to ask whether black holes and AGN do also
 show a similar trend. The first hints of a positive answer came from
 the study of the evolution of the X-ray selected AGN luminosity
 function. As X-rays are produced very deep in the gravitational
 potential of a black hole, they are the best probes of the accretion
 energy release (modulo the effect of obscuration). 
In the last decade, we have thus learned 
that more luminous AGN were more common in
 the past, with the X-ray luminosity function (XLF) following a
 so-called 
'Luminosity Density Dependent Evolution' \citep{ueda:03,hasinger:05}. 
These works were all based on a combination of a large number
 of X-ray surveys of different depth and area. Now the XMM-COSMOS
 survey \citep{hasinger:07} {\it alone}, thanks to its very high
 spectroscopic completeness over a relatively large area of the sky,
 can be used to provide new constraints on the XLF, especially at
 $z>3$ \citep{brusa:09}.
The left panel of Figure~\ref{fig:xmm} (Miyaji et
al. in prep.) shows the number density evolution of X-ray selected AGN 
 of different luminosities
obtained from COSMOS data {\it only}, where one can clearly see how
more luminous AGN were more common in the past as compared to
lower luminosity ones, a direct phenomenological manifestation of AGN
downsizing. 
How can we use this (and other analogous) results on the XLF evolution
to gain further insights on the physical evolution of the black hole population?

As opposed to the case of galaxies, where the direct relationship
between the evolving mass functions of the various morphological types
and the distribution of star forming galaxies is not straightforward
due to the never-ending morphological and photometric transformation
of the different populations, the situation in the case of SMBH is
much simpler. For the latter case, we can assume their evolution is
governed by a continuity equation (MH08, and references therein), 
where the mass function of SMBH at any given time can be used
to predict that at any other time, provided the distribution of
accretion rates as a function of black hole mass is known. 
Such equation  can be written as:
\begin{equation}
\label{eq:continuity}
\frac{\partial \psi(\mu,t)}{\partial t} +
\frac{\partial}{\partial \mu}\left( \psi(\mu,t) \langle \dot M
  (\mu,t)\rangle \right)=0
\end{equation}
where  $\mu=Log\, M$ ($M$ is the black hole mass in solar units),
$\psi(\mu,t)$ is the SMBH mass function at time $t$, and $\langle \dot M
(\mu,t) \rangle$ is the average accretion rate of SMBH of mass $M$ at
time $t$, and can be defined through a ``fueling'' function,
$F(\dot\mu,\mu,t)$, describing the
distribution of accretion rates for objects of mass $M$ at time $t$:
$\langle \dot M(M,z)\rangle = \int \dot M F(\dot\mu,\mu,z)\, \mathrm{d}\dot\mu$.
Such a fueling function is not a priori known, and observational
determinations thereof have been able so far to probe robustly only
the extremes of the overall population.   
However, the AGN fueling function 
can be derived by inverting the integral
equation that relates the luminosity function of the population in
question with its mass function. Indeed we can write:
\begin{equation}
\label{eq:filter}
\phi(\ell,t)=\int F(\dot\mu,\mu,t) \psi(\mu,t)\; \mathrm{d}\mu
\end{equation}
where I have called $\ell=Log\, L_{\rm bol}$.
This is the approach followed in MH08, were the inversion was
performed numerically,
based on a minimization scheme that used both the X-ray and radio AGN
luminosity functions as constraints, complemented by recipes to relate
observed (and intrinsic) X-ray and radio (core) luminosities to
$L_{\rm bol}$ (see MH08 for details).

Using this approach, 
we have integrated eq~(\ref{eq:continuity}) starting from $z=0$, where
we have simultaneous knowledge of both mass, $\psi(\mu)$, and
luminosity, $\phi(\ell)$, functions,
evolving the SMBH mass function backwards in time, up to where
reliable estimates of the (hard X-ray selected) 
AGN luminosity functions are available (currently this means $z\simeq 4$).
The adopted hard X-ray luminosity function is
supplemented with luminosity-dependent bolometric
corrections of \citet{marconi:04}
and absorbing column density distributions 
consistent with the X-ray background  constraints, following
the most recent XRB synthesis model (see \citet{gilli:07} for
details). Similar results can of course be obtained using directly
bolometric luminosity functions (see e.g. \citet{hopkins:07} for a
recent attempt to determine $\phi(\ell,t)$).

In this way, we can estimate the specific
instantaneous ratio of black hole mass to accretion rate as a function
of SMBH mass and its cosmological evolution. 
Such a ratio defines a timescale, the so-called {\it growth
time}, or mass doubling time (Figure~\ref{fig:xmm}, right). 
 The redshift evolution of the growth time distribution
can be used to identify the epochs when black holes of different sizes
grew the largest fraction of their mass: Black holes
with growth times longer than the age of the Universe are not
experiencing a major growth phase, which must have necessarily
happened at earlier times. Figure~\ref{fig:xmm} then shows that, while
at $z<0.5-1$ only black holes with masses smaller than $10^7
M_{\odot}$ are experiencing significant growth, as
we approach the peak of the black hole accretion rate density ($z \sim
1.5-2$), we witness the rapid growth of the entire SMBH
population. Better constraints on both bolometric luminosity and mass
functions evolution are however needed to paint a clearer picture at
higher $z$.

\subsection{The evolution of scaling relations}
Modern multiwavelength surveys are increasingly 
designed to allow measurements of the
physical properties of AGN hosts. 
Within COSMOS, we have recently studied the hosts of 89 broad line (type--1) Active Galactic Nuclei (AGN)  detected
in the zCOSMOS survey \citep{lilly:07} in the
redshift range $1<z<2.2$ (for all the details, see \citet{merloni:09}; M09). 
The unprecedented multi-wavelength coverage of
the survey field allowed us to
disentangle the emission of the host galaxy from that of the nuclear
black hole in their Spectral Energy Distributions (SED).
We derive an estimate of black hole masses
through the analysis of the broad MgII emission
lines observed in the
medium-resolution spectra taken with {\it VIMOS/VLT} as part of the
zCOSMOS project. Then, we estimated rest frame K-band luminosity and
total stellar 
mass (and their corresponding uncertainties) of the AGN hosts through
an extensive SED fitting procedure, based on large databases of both
phenomenological and theoretical galaxy spectra.

\begin{figure}
  \includegraphics[width=10cm,height=.30\textheight]{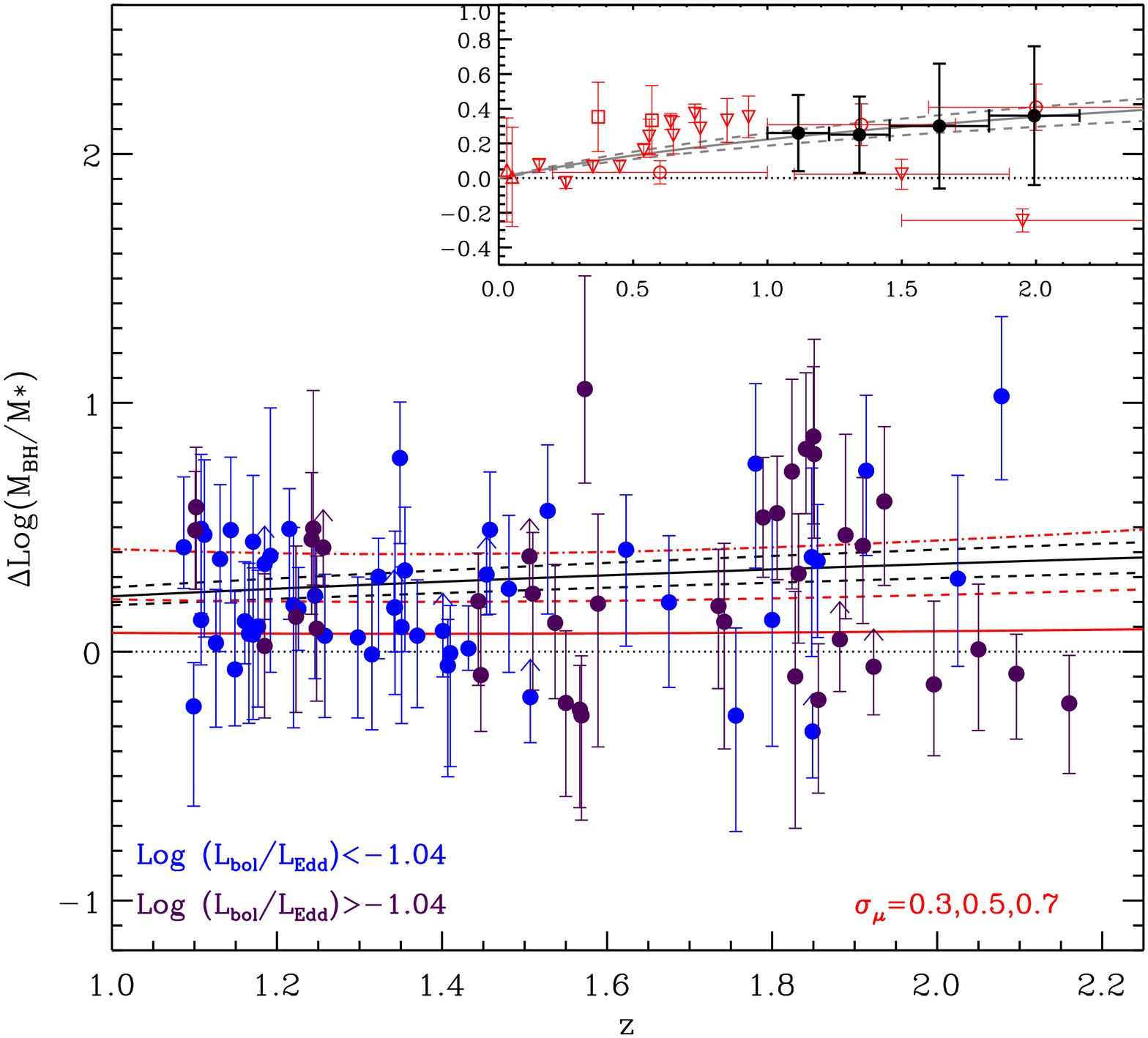}
  \caption{Redshift evolution of the offset measured for our type--1 AGN from
  the local $M_{\rm BH}-M_{*}$ relation. Different colors identify different
ranges of Eddington ratios with upwards arrows representing upper limits on 
the host mass. The offset is calculated as the
distance of each point to the \citet{haering:04} correlation. Solid black
line shows the best fit obtained assuming an evolution of the form
$\Delta {\rm Log}(M_{\rm BH}/M_{*})(z)=\delta_2{\rm Log}(1+z)$; for which we
found $\delta_2=0.74 \pm 0.12$. The red lines show the bias due to the intrinsic
scatter in the scaling relation to be expected even if they are universal. Solid line
is for an intrinsic scatter of 0.3 dex; dashed of 0.5 dex; dot-dashed
of 0.7 dex. The inset shows a comparison with literature results. 
From \citet{merloni:09}}
   \label{fig:dmbh_z}
\end{figure}

We found that, as compared to the local value, the
average black hole to host galaxy mass ratio appears to evolve positively with
redshift (see Figure~\ref{fig:dmbh_z}), with a best fit evolution of the form
$(1+z)^{0.74 \pm 0.12 ^{+0.6}_{-0.3}}$, where the large asymmetric systematic
errors stem from the uncertainties in the choice of IMF, in the
calibration of the virial relation used to estimate BH masses and in
the mean QSO SED adopted. 
A thorough analysis of observational biases induced by
intrinsic scatter in the scaling relations reinforces the conclusion
that an evolution of the $M_{\rm BH}-M_{*}$ relation must
ensue for actively growing black holes at early times:
either its overall normalization, or its intrinsic scatter (or
both) appear to increase with redshift. 
This can be interpreted as signature of either a more rapid growth of supermassive
black holes at high redshift, or a significant mismatch between the typical growth
times of nuclear black holes and host galaxies. In both cases, our results
provide important clues on the nature of the early co-evolution of
black holes and galaxies and challenging tests
for models of AGN feedback and self-regulated growth of structures.



\begin{theacknowledgments}
I am grateful to my collaborators A. Bongiorno, M. Brusa, S. Heinz,
T. Miyaji and all the members of the COSMOS and zCOSMOS teams for their 
essential contribution to the work presented here. 
\end{theacknowledgments}



\bibliographystyle{aipproc}   

\bibliography{merloni_a}

\IfFileExists{\jobname.bbl}{}
 {\typeout{}
  \typeout{******************************************}
  \typeout{** Please run "bibtex \jobname" to optain}
  \typeout{** the bibliography and then re-run LaTeX}
  \typeout{** twice to fix the references!}
  \typeout{******************************************}
  \typeout{}
 }

\end{document}